\documentclass{ws-mpla}
\pdfoutput=1
\usepackage[T1]{fontenc}
\usepackage{bm,amsmath,amssymb,color,bbm,mathrsfs,booktabs,makecell}
\usepackage[super]{cite}
\usepackage{xcolor}
\usepackage[verbose,hypertexnames=false]{hyperref}
\hypersetup{colorlinks=false,allbordercolors=blue,pdfborderstyle={/S/U/W 1}}
\usepackage{slashed}

\DeclareMathOperator\diag{diag}
\DeclareMathOperator\Tr{Tr}

\DeclareMathOperator\rme{\mathrm{e}}

\renewcommand{\bar}[1]{\overline{#1}}

\newcommand{\bep}{\begin{pmatrix}} 
\newcommand{\eep}{\end{pmatrix}}

\newcommand{\SU}{\text{SU}}

\newcommand{\U}{\text{U}}
\newcommand{\1}{\mathbbm{1}}
\newcommand{\RR}{\mathbb{R}}
\newcommand{\CC}{\mathbb{C}}

\renewcommand{\epsilon}{\varepsilon}
\newcommand{\rmd}{\mathrm{d}}

\newcommand{\up}{\uparrow}
\newcommand{\down}{\downarrow}
\newcommand{\m}{\widehat{m}}
\renewcommand{\j}{\widehat{j}}
\newcommand{\J}{\widehat{J}}

\def\ba#1\ea{\begin{align}#1\end{align}}
\def\mkakko#1{\left( #1 \right)}
\def\ckakko#1{\left\{ #1 \right\}}
\def\kkakko#1{\left[ #1 \right]}
\def\wt#1{\widetilde{#1}}
\def\revision#1{{\color{black}#1}}

\begin{document}

\markboth{T.~Kanazawa}{Analysis of the QCD Kondo phase using random matrices}

\catchline{}{}{}{}{}

\title{Analysis of the QCD Kondo phase using random matrices}

\author{Takuya Kanazawa}
\address{Department of Business Administration, Kobe Gakuin University, 1-1-3 Minatojima, Chuo-ku, Kobe 650-8586, Japan
\\
tkanazawa@ba.kobegakuin.ac.jp}

\allowdisplaybreaks
\maketitle

\begin{history}
\received{(Day Month Year)}
\revised{(Day Month Year)}
\accepted{(Day Month Year)}
\published{(Day Month Year)}
\end{history}

\begin{abstract}
We propose a novel random matrix model that describes the QCD Kondo phase. The model correctly implements both the chiral symmetry of light quarks and the $\SU(2)$ spin symmetry of heavy quarks. We analytically take the large-$N$ limit with $N$ the matrix size and show that the model has three phases: the pure Kondo phase with no chiral condensate, the pure chirally broken phase with no Kondo condensate, and the coexistence phase. The model predicts that the pairing form of the Kondo condensate in the coexistence phase is significantly altered compared to the pure Kondo phase. For each phase, we rigorously derive the low-energy effective theory of Nambu-Goldstone modes and obtain compact closed expressions for the partition function with external sources. We also include a chiral chemical potential into the model and examine the vacuum structure.  
\end{abstract}


\ccode{PACS Nos.: 02.10.Yn, 12.39.Fe, 21.65.Qr}

\section{Introduction}

In certain alloys with dilute magnetic impurities, the electrical resistivity shows a minimum at nonzero temperature, a phenomenon famously known as the Kondo effect \cite{Kondo:1964nea,Hewson_1993,Coleman2006,Hewson:2009}. The antiferromagnetic coupling between conduction electrons and an impurity gets stronger as the temperature lowers, and eventually leads to a complete screening of the magnetic moment of the impurity. This is a classical example of asymptotic freedom. The prerequisites of the Kondo effect are (i) nonzero density of states of conduction electrons, (ii) a non-Abelian interaction, and (iii) quantum loop effects. Later the Kondo problem was investigated for materials which have the density of states that vanishes at the Fermi energy as a power law, $\rho(E)\propto|E-E_F|^r$. This is called the pseudogap Kondo problem and is relevant to materials hosting quasiparticles with relativistic dispersions, including graphene, Dirac/Weyl semimetals, and $d$-wave superconductors \cite{PhysRevLett.64.1835,PhysRevB.53.15079,PhysRevB.57.14254,PhysRevB.70.214427,PhysRevB.74.144410,ChenNature2011,Fritz_2013,PhysRevB.92.041107,PhysRevB.92.121109,PhysRevB.92.195124,doi:10.7566/JPSJ.84.074705,PhysRevB.98.075110}. In these systems, there is a quantum phase transition at finite coupling from the unscreened free moment phase to the strongly coupled Kondo phase. This is analogous to chiral symmetry breaking in the QCD vacuum \cite{Nambu:1961tp,Nambu:1961fr}.  

Recently the Kondo effect in relativistic nuclear and quark matter has been discussed \cite{Yasui:2013xr,Hattori:2015hka,Ozaki:2015sya,Yasui:2016ngy,Yasui:2016svc,Kanazawa:2016ihl,Yasui:2017izi,Suzuki:2017gde,Kimura:2018vxj,Hattori:2019zig,Fariello:2019ovo,Suenaga:2019jqu,Suenaga:2019car}. In such environments the impurity is either a single heavy quark (i.e., charm and bottom quarks) or a hadron including a heavy quark (e.g., $D$ and $B$ mesons). The non-Abelian symmetry pertinent to the QCD Kondo effect is the $\SU(2)$ isospin symmetry and the $\SU(3)$ color symmetry. To be precise, as first pointed out in Ref.~\refcite{Kanazawa:2016ihl}, the Kondo effect in QCD is a relativistic analog of the \emph{overscreened Kondo effect} \cite{Nozieres1980,Affleck:1990zd,Affleck:1990iv,Parcollet:1997ysb}, in which the low-energy physics is governed by a non-Fermi-liquid fixed point. The Kondo scale in quark matter is estimated to be $10\sim100$ MeV \cite{Yasui:2016svc,Suzuki:2017gde} while that in normal metals is of order 10~K, so they are different by a factor of $10^{11}$. It is quite impressive that the same physical mechanism is operative at such drastically different scales. Since heavy flavors are phenomenologically important as a probe of hot and dense matter in heavy ion collisions and compact stars \cite{Neubert:1993mb,Neubert:1996wg,Manohar_Wise_2023,Hosaka:2016ypm}, we expect the QCD Kondo effect to play a role there. However it must be noted that the QCD Kondo effect is always subject to competitions with other kind of condensates \cite{Kanazawa:2016ihl,Suzuki:2017gde}. At low baryon density the chiral condensate is dominant, while at asymptotically high baryon density the diquark condensate is dominant.\cite{Rajagopal:2000wf,Alford:2007xm,Fukushima:2010bq} Therefore in these two regions the QCD Kondo effect could be strongly suppressed. Studies \cite{Hattori:2015hka,Ozaki:2015sya,Suenaga:2019jqu} have shown that a baryon chemical potential, an external magnetic field, and a chiral chemical potential all serve as a catalyst of the QCD Kondo effect; however, at the same time they also catalyze chiral symmetry breaking and color superconductivity, so the fate of the QCD Kondo effect is, to say the least, elusive. Nevertheless it is interesting in its own right as a crossroad of condensed matter physics and high energy physics. 

Random matrix theory (RMT) is a powerful framework to analyze statistical properties of interacting many-body systems \cite{Guhr:1997ve,Akemann:2011csh}. RMT has been applied to QCD to derive universal microscopic correlation functions of the Dirac eigenvalues at zero \cite{Shuryak:1992pi,Verbaarschot:1993pm,Verbaarschot:1994ia,Akemann:1996vr,Damgaard:1997ye} and nonzero chemical potential \cite{Osborn:2004rf,Akemann:2004dr,Osborn:2005ss,Akemann:2005fd,Kanazawa:2009en,Akemann:2010tv,Kanazawa:2011tt,Kanazawa:2014lga,Kanazawa:2020dju,Kanazawa:2020ktn,Kanazawa:2020fpo}, as well as to map out the QCD phase diagram \cite{Halasz:1998qr,Vanderheyden:2000ti,Pepin:2000pv,Vanderheyden:2001gx,Klein:2003fy,Klein:2004hv,Vanderheyden:2011iq} (see \cite{Verbaarschot:1997bf,Verbaarschot:2000dy,Verbaarschot:2005rj,Akemann:2007rf,Kanazawabook} for reviews). The matching between RMT and lattice simulation data provides a clean way to determine low-energy constants in chiral perturbation theory \cite{BerbenniBitsch:1997tx,Fukaya:2007fb}. Furthermore, since RMT is analytically tractable and yet captures the essence of QCD, it has long been used as a testing ground of newly developed methods to solve the sign problem in QCD, such as the complex Langevin algorithm \cite{Mollgaard:2013qra,Nagata:2016alq,Bloch:2017sex}, the subset method \cite{Bloch:2012ye}, and the Lefschetz thimbles \cite{DiRenzo:2015foa}.  

In this paper, we aim to build the first random matrix model for the QCD Kondo effect. On the basis of symmetry principles, we construct the thermodynamic potential, find the minimum, and clarify the phase diagram, in an analytically controlled manner. In contrast to the preceding works \cite{Hattori:2015hka,Kanazawa:2016ihl,Yasui:2017izi,Suzuki:2017gde,Suenaga:2019car}, this paper specifically elucidates the impact of (nearly) gapless Nambu-Goldstone (NG) modes stemming from spontaneous symmetry breaking. Such modes have been dismissed in the conventional mean-field approximation because the direction of the condensate is fixed in the internal symmetry space. We show how to integrate out the soft fluctuations around the ground state nonperturbatively. Summarizing, this work opens up a novel way to study quark matter with heavy flavors using the simple yet versatile framework of RMT.

This paper is organized as follows. In Section~\ref{sc:ewsf} we sort out the patterns of symmetry breaking and give an intuitive picture of the QCD Kondo phase. In Section~\ref{sc:mm} we define the new random matrix model and clarify symmetries. In Section~\ref{sc:phases} the phase structure of the model for $N\gg 1$ is investigated, where $N$ denotes the matrix size. In Section~\ref{sc:eft} we add external source terms and derive the effective theory of NG modes. In Section~\ref{sc:ccp} we introduce a chiral chemical potential into the model and study the behavior of Kondo condensates. In Section~\ref{sc:conc} we conclude.

\section{Symmetries in the QCD Kondo phase}\label{sc:ewsf}

Since the guiding principle of RMT is the correct matching of symmetries, one must first understand symmetries of the QCD Kondo phase. Let us consider heavy quarks $Q$ that are uniformly distributed in the medium of $N_f$ light flavors $\psi$. As usual, there is a global symmetry $\U(1)_{\rm V}\times\U(1)_{\rm A}\times\SU(N_f)_{\rm L}\times\SU(N_f)_{\rm R}$ for $\psi$, under the assumption that current quark masses and the charge neutrality condition can be ignored. By contrast, it makes no sense to talk about chiral symmetry for $Q$. Instead, the relevant symmetry of $Q$ is $\U(1)_{\rm Q} \times\SU(2)_{\rm H}$, which is called the heavy quark symmetry (HQS) \cite{Neubert:1993mb,Neubert:1996wg,Manohar_Wise_2023,Hosaka:2016ypm}.\footnote{In the presence of $N_h$ flavors of heavy quarks, the heavy quark symmetry is enlarged to $\U(1)_{\rm Q}\times\SU(2N_h)$. However, in this paper we only consider a single heavy flavor for simplicity of exposition.}  This $\SU(2)_{\rm H}$ symmetry originates from the fact that when the mass $m_Q$ of $Q$ is large, interactions that flip the spin of $Q$ are suppressed by $1/m_Q$. Now, according to the putative QCD Kondo effect, the heavy-light condensate $\langle \bar\psi Q \rangle$ forms%
\footnote{This is a relativistic analog of the hybridization of impurity spin and conduction electrons in the Coqblin-Schreiffer model \cite{Read_1983,Bickers:1987zz}.} and triggers spontaneous symmetry breaking \cite{Yasui:2017izi,Yasui:2016svc}
\begin{gather}
	\U(1)_{\rm V}\times\U(1)_{\rm A}\times\SU(N_f)_{\rm L}\times\SU(N_f)_{\rm R} \times \U(1)_{\rm Q} \times\SU(2)_{\rm H} 
	\notag
	\\
	\to \U(1)_{\rm V+Q} \times \U(1)_{\rm A+H} \times \SU(N_f-1)_{\rm L} \times \SU(N_f-1)_{\rm R}\,. 
	\label{eq:Ksym}
\end{gather}
Here we of course assumed that there is no other condensate that interferes with the Kondo effect. In \eqref{eq:Ksym}, $\U(1)_{\rm A+H}$ represents the dynamical locking of $\U(1)_{\rm A}$ and an Abelian subgroup of $\SU(2)_{\rm H}$. It is dubbed as \emph{chiral-HQS locking} \cite{Yasui:2017izi,Yasui:2016svc} and is a hallmark of the QCD Kondo effect. 

In this paper, we aim to construct RMT that correctly implements both the chiral symmetry of light quarks and the heavy quark symmetry of impurities. By taking the large-$N$ limit where $N$ is the matrix size, we will later show for $N_f=1$ that there appear three distinct phases%
\footnote{Since our RMT is effectively zero-dimensional and, in the large-$N$ limit, is not subject to spatial (infrared) fluctuations that could restore the symmetry, the symmetry-unbroken phase does not appear in its phase diagram.}:
\begin{itemize}
	\item Pure Kondo phase:~$\langle\bar\psi Q\rangle\ne 0$ and $\langle\bar\psi\psi\rangle=0$.
	\item Pure chirally broken phase:~$\langle\bar\psi Q\rangle=0$ and $\langle\bar\psi\psi\rangle\ne0$.
	\item Coexistence phase:~$\langle\bar\psi Q\rangle\ne 0 \ne \langle\bar\psi\psi\rangle$.
\end{itemize}
One can go to each phase by deliberately tuning a parameter of the model that controls the relative strength of interactions for $\bar\psi\psi$ and $\bar\psi Q$. For the pure Kondo phase, we show that the symmetry breaking pattern coincides with \eqref{eq:Ksym}. In particular, the chiral-HQS locking is faithfully realized. For the coexistence phase, we reveal that the pairing form of the Kondo condensate is drastically modified due to the influence of the chiral condensate, which has not been known before. We will also include the chiral chemical potential and examine how the Kondo condensate is deformed.

\section{\label{sc:mm}Random matrix partition function}

\revision{\noindent
The RMT we propose is defined by the partition function
\ba
	Z & = \int \rmd \wt{W} \int \rmd \wt{V} 
	~\rme^{-N \Tr(\wt{W}^\dagger\wt{W})-N\Tr(\wt{V}^\dagger \wt{V})}
	\notag
	\\
	& \quad \times \prod_{f=1}^{N_f}
	\det(\wt{\mathscr{D}}
	+ \wt{m}_f\1_{2N})\times 
	{\det}^2\big(- g_{Q} \wt{V}^\dagger + \wt{\lambda} \1_N \big)\,,\label{eq:t4dv87re}
\ea
where $\wt{W}$ and $\wt{V}$ are $N\times N$ complex matrices. The Dirac operator for light quarks is given by the non-Hermitian $2N\times 2N$ matrix
\ba
	\wt{\mathscr{D}}\equiv \begin{pmatrix}
	\mathbf{0}_N & g_q \wt{W} + g_{Q} \wt{V}
	\\
	- g_q \wt{W}^\dagger + g_{Q} \wt{V} & \mathbf{0}_N
	\end{pmatrix}\,.
	\label{eq:qpmxoejrg}
\ea
In \eqref{eq:t4dv87re} and \eqref{eq:qpmxoejrg}, $g_q$ and $g_{Q}$ represent the interaction strength between light-light and light-heavy quarks, respectively. Let us expel powers of $g_Q$ from the determinants
\ba
	& \prod_{f=1}^{N_f}
	\det(\wt{\mathscr{D}}
	+ \wt{m}_f\1_{2N})\times 
	{\det}^2\big(- g_{Q} \wt{V}^\dagger 
	+ \wt{\lambda} \1_N \big)
	\notag
	\\
	& \propto \prod_{f=1}^{N_f}
	\det \mkakko{\frac{\wt{\mathscr{D}}}{g_Q}
	+ \frac{\wt{m}_f}{g_Q}\1_{2N}}
	\times 
	{\det}^2 \mkakko{- \wt{V}^\dagger 
	+ \frac{\wt{\lambda}}{g_{Q}} \1_N }
\ea
with
\ba
	\frac{\wt{\mathscr{D}}}{g_Q} = 
	\begin{pmatrix}
	\mathbf{0}_N & \frac{g_q}{g_Q} \wt{W} + \wt{V}
	\\
	- \frac{g_q}{g_Q} \wt{W}^\dagger + \wt{V} & \mathbf{0}_N
	\end{pmatrix},
\ea
and then define new variables as
\ba
\begin{array}{c}
	\displaystyle 
	W \equiv \frac{g_q}{g_Q} \wt{W}, 
	\quad 
	V \equiv \wt{V}, 
	\quad
	m_f \equiv \frac{\wt{m}_f}{g_Q},
	\vspace{2mm}\\
	\displaystyle 
	\mathscr{D} \equiv \frac{\wt{\mathscr{D}}}{g_Q}
	= \begin{pmatrix}
	  \mathbf{0}_N & W + V
	  \\
	  -W^\dagger + V & \mathbf{0}_N
	\end{pmatrix},
	\quad
	\lambda \equiv \frac{\wt{\lambda}}{g_{Q}},
	\quad \eta \equiv \frac{g_Q^2}{g_q^2}.
\end{array}
	\label{eq:2YYYRE}
\ea
This leads us to the slightly simpler form of the partition function
\ba
	Z & = \int \rmd W \int \rmd V ~
	\rme^{-N\eta\Tr(W^\dagger W)-N\Tr(V^\dagger V)}
	\notag
	\\
	& \quad \times \prod_{f=1}^{N_f}\det(\mathscr{D}+m_f\1_{2N})\times {\det}^2(-V^\dagger + \lambda \1_N)\,.
	\label{eq:ZRMT}
\ea 
}

In the following, we focus on $N_f=1$ in the chiral limit for technical simplicity. Let us introduce light quark fields $\psi_{i\alpha}$ and $\bar\psi_{i\alpha}$ ($i={\rm R,L}$) as well as heavy quark fields with spin $Q_{\up\alpha}, Q_{\down\alpha}, \bar Q_{\up\alpha}, \bar Q_{\down\alpha}$ where $\alpha$ runs from 1 to $N$.\footnote{We suppress the spinor index of $\psi$ because it is not needed for the description of symmetry breaking.} With the aid of these Grassmann variables, we obtain
\ba
	Z & = \int \rmd \bar\psi\, \rmd \psi \,\rmd \bar Q\, \rmd Q 
	\int \rmd W \int \rmd V 
	\exp\big[-N\eta\Tr(W^\dagger W)-N\Tr(V^\dagger V)
	\notag
	\\
	& \quad +\bar\psi_{{\rm R}\alpha}(W+V)_{\alpha\beta}\psi_{{\rm R}\beta}
	+\bar\psi_{{\rm L}\alpha}(-W^\dagger+V)_{\alpha\beta}\psi_{{\rm L}\beta}
	\notag
	\\
	& \quad + \bar{Q}_{\up\alpha}(-V^\dagger + \lambda \1_N)_{\alpha\beta}Q_{\up\beta}
	+ \bar{Q}_{\down\alpha}(-V^\dagger + \lambda \1_N)_{\alpha\beta}Q_{\down\beta}
	\big]
\ea
Clearly this model has the symmetry $\U(1)_{\rm V}\times\U(1)_{\rm A}$ for $\psi$ and $\U(1)_{\rm Q}\times\SU(2)_{\rm H}$ for $Q$. 

Let us integrate out the Gaussian matrices and perform the Hubbard-Stratonovich transformation to make taking the large-$N$ limit easier. For the $W$ part, we have
\ba
	& \int \rmd W \exp(
	-N\eta W^*_{\alpha\beta}W_{\alpha\beta}+\bar\psi_{{\rm R}\alpha}W_{\alpha\beta}\psi_{{\rm R}\beta}
	-\bar\psi_{{\rm L}\beta}W^*_{\alpha\beta}\psi_{{\rm L}\alpha}
	)
	\notag
	\\
	& \propto \exp\mkakko{-\frac{1}{N\eta}\bar\psi_{{\rm R}\alpha}\psi_{{\rm R}\beta}\bar\psi_{{\rm L}\beta}\psi_{{\rm L}\alpha}}
	\\
	& \propto \int_{\CC}\rmd \sigma~\exp(
		- N\eta|\sigma|^2 + \bar\psi_{{\rm R}\alpha}\psi_{{\rm L}\alpha}\sigma + \bar\psi_{{\rm L}\alpha}\psi_{{\rm R}\alpha}\sigma^*
	)\,.
\ea
The $V$ part reads
\ba
	& \int \rmd V \exp \big[
		-N V^*_{\alpha\beta}V_{\alpha\beta} 
		+ (\bar\psi_{{\rm R}\alpha}\psi_{{\rm R}\beta} + \bar\psi_{{\rm L}\alpha}\psi_{{\rm L}\beta})V_{\alpha\beta}
		- (\bar{Q}_{\up\beta}Q_{\up\alpha}+\bar{Q}_{\down\beta}Q_{\down\alpha}) V^*_{\alpha\beta}
	\big]
	\notag
	\\
	& \propto \exp\kkakko{- \frac{1}{N}(\bar\psi_{{\rm R}\alpha}\psi_{{\rm R}\beta}+\bar\psi_{{\rm L}\alpha}\psi_{{\rm L}\beta})
	(\bar{Q}_{\up\beta}Q_{\up\alpha}	+ \bar{Q}_{\down\beta}Q_{\down\alpha})}
	\\
	& \propto \int_{\CC^4}\rmd K~\exp \big[
	-N\Tr(K^\dagger K)+\bar\psi_{{\rm R}\alpha}K_{{\rm R}\up}Q_{\up\alpha}
	+ \bar{Q}_{\up\alpha}K^*_{{\rm R}\up}\psi_{{\rm R}\alpha}
	+ \bar\psi_{{\rm L}\alpha}K_{{\rm L}\up}Q_{\up\alpha} 
	\notag
	\\
	& \quad 
	+ \bar{Q}_{\up\alpha}K^*_{{\rm L}\up}\psi_{{\rm L}\alpha}
	+\bar\psi_{{\rm R}\alpha}K_{{\rm R}\down}Q_{\down\alpha}
	+ \bar{Q}_{\down\alpha}K^*_{{\rm R}\down}\psi_{{\rm R}\alpha}
	+ \bar\psi_{{\rm L}\alpha}K_{{\rm L}\down}Q_{\down\alpha}
	+ \bar{Q}_{\down\alpha}K^*_{{\rm L}\down}\psi_{{\rm L}\alpha}
	\big]\,,
\ea
where the notation $K\equiv \begin{pmatrix}K_{{\rm R}\up}&K_{{\rm R}\down}\\K_{{\rm L}\up}&K_{{\rm L}\down}\end{pmatrix}$ was used. Collecting everything, we obtain
\ba
	Z & \propto \int \rmd\bar\psi\, \rmd \psi\,
	\rmd \bar{Q}\, \rmd Q \int_\CC \rmd \sigma \int_{\CC^4}\rmd K ~
	\rme^{- N\eta|\sigma|^2 -N\Tr(K^\dagger K)}
	\notag
	\\
	& \qquad \times \exp\left[
	\begin{pmatrix}\bar\psi_{\rm R}\\\bar\psi_{\rm L}\\\bar{Q}_\up\\\bar{Q}_\down\end{pmatrix}_\alpha
	\hspace{-4pt}
	\left(\begin{array}{c|c}
	\begin{matrix}0&\sigma\\\sigma^*&0\end{matrix}& K
	\\\hline 
	K^\dagger &\begin{matrix}\lambda&0\\0&\lambda\end{matrix}
	\end{array}\right)
	\begin{pmatrix}\psi_{\rm R}\\\psi_{\rm L}\\Q_\up\\Q_\down\end{pmatrix}_\alpha
	\right]
	\label{eq:rq354ewd}
	\\
	& \propto \int_\CC \rmd \sigma \int_{\CC^4}\rmd K~
	\rme^{- N\eta|\sigma|^2 -N\Tr(K^\dagger K)}\;
	{\det}^N \left[
		KK^\dagger - \lambda \begin{pmatrix}0&\sigma\\\sigma^*&0\end{pmatrix}
	\right]
	\\
	& = \int_\CC \rmd \sigma \int_{\CC^4}\rmd K~z(\sigma,K)^N,
	\label{eq:qghgWWWW1}
\ea
where we have defined (with rescaling $\sigma\to \sigma/\sqrt{\eta}$)
\ba
	z(\sigma,K) \equiv \rme^{-|\sigma|^2-\Tr(K^\dagger K)}
	\det \left[
		KK^\dagger - \xi \begin{pmatrix}0&\sigma\\\sigma^*&0\end{pmatrix}
	\right]
	\label{eq:smz}
\ea
with \revision{
\ba
	\xi \equiv \frac{\lambda}{\sqrt{\eta}} 
	= \lambda \frac{\left| g_q \right|}{\left| g_Q \right|} \,.
	\label{eq:xxiidef}
\ea
In the last step, the definition of $\eta$ \eqref{eq:2YYYRE} has been substituted. The parameter $\xi$ controls the ratio of light-light to light-heavy quark interactions.} Note that \eqref{eq:qghgWWWW1} and \eqref{eq:smz} are exact rewriting of the original partition function and so far no approximation has been made. The phase diagram of the model as a function of $\xi$ is the subject of the next section.

\section{\boldmath Phase structure at large $N$\label{sc:phases}}

\begin{figure}[tb]
	\centering
	\includegraphics[width=0.7\columnwidth]{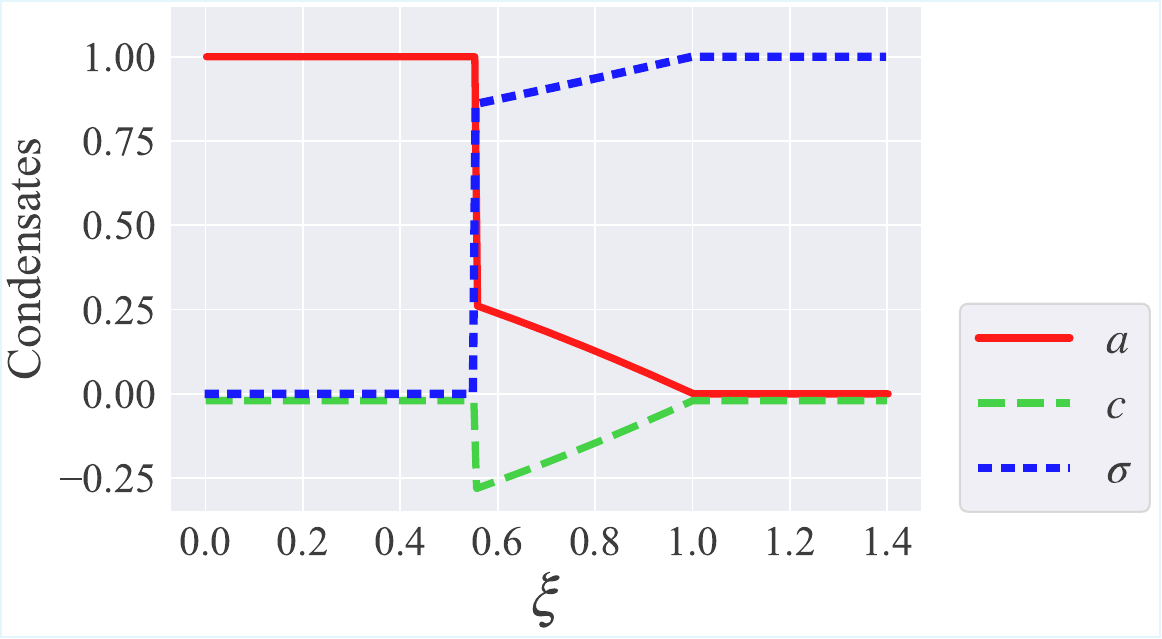}
	\caption{\label{fg:acs}\revision{The condensates at the minimum of the free energy \eqref{eq:fre} are plotted as a function of $\xi$. ($b$ is not plotted because $a=b$ for all $\xi$.) 
	Here, $a$ and $c$ characterize the Kondo condensate, while $\sigma$ corresponds to the chiral condensate.}}
\end{figure}

For $N\gg 1$, the ground state of the model is determined from the saddle point analysis of the function $z(\sigma,K)$ in \eqref{eq:smz}. Through a $\U(1)_{\rm A}$ rotation $K\to \exp(i\theta \sigma_3)K$ the phase of $\sigma$ can be arbitrarily tuned, so one can assume $\sigma \geq 0$ without loss of generality. For numerics, we adopt the parameterization
\ba
	KK^\dagger = \begin{pmatrix}a&c+id\\c-id&b\end{pmatrix}
	\label{eq:KKp}
\ea
with
\ba
	a,b,c,d\in\RR, \quad a\geq 0,~b\geq 0,~ab\geq c^2+d^2\,.
\ea
Then
\ba
	z(\sigma,K) & = \rme^{-\sigma^2-a-b}\det
	\begin{pmatrix}
		a&c+id-\xi \sigma \\ c-id-\xi \sigma &b
	\end{pmatrix}
	\\
	& = \rme^{-\sigma^2-a-b}[ab-(c-\xi\sigma)^2-d^2] \,.
\ea
Suppose we rotate the vector $(c,d)\in\RR^2$ with $\sigma,a,b$ and $c^2+d^2$ fixed. Since $ab-(c-\xi\sigma)^2-d^2=ab-c^2-d^2-\xi^2\sigma^2+2c\xi\sigma=2\begin{pmatrix}c\\d\end{pmatrix}\cdot\begin{pmatrix}\xi\sigma\\0\end{pmatrix}+\text{const.}$, it is obvious that $ab-(c-\xi\sigma)^2-d^2$ is extremized if and only if the vector $(c,d)$ is parallel to $(\xi\sigma,0)$. Thus $d=0$ at the saddle point. Taking this observation into account, we find that the task is to minimize the free energy
\ba
	f(a,b,c,\sigma) & \equiv \sigma^2 + a + b - \log \left|ab-(c-\xi\sigma)^2 \right|
	\label{eq:fre}
\ea
under the conditions $\sigma\geq0,a\geq 0, b\geq0$ and $ab\geq c^2$. We numerically solved this problem. The result is plotted in Figure~\ref{fg:acs}. 
A summary of the phase structure is given in the table below. 
\begin{center}
	\begin{tabular}{@{}ccc@{}}
		\toprule
		Domain & Phase & Condensates 
		\\
		\midrule \midrule
		$0 \le \xi < 0.556$
		& \makecell{Pure Kondo\\phase}
		& $\langle K\rangle=\1_{2},\ \langle\sigma\rangle=0$ 
		\\
		\midrule
		$0.556 \le \xi \le 1$
		& \makecell{Coexistence\\phase}
		& $\langle K\rangle\neq 0,\ \langle\sigma\rangle\neq 0$ 
		\\
		\midrule
		$\xi > 1$
		& \makecell{Pure chirally\\broken phase}
		& $\langle K\rangle=0,\ \langle\sigma\rangle=1$ \\
		\bottomrule
	\end{tabular}
\end{center}

\revision{Recalling \eqref{eq:xxiidef}, the overall trend is physically intuitive: the Kondo effect dominates when the heavy-light interaction is stronger (small $\xi$). Conversely, chiral symmetry breaking dominates when the light-light interaction is stronger (large $\xi$). We discuss this in more detail below.}

In the pure Kondo phase at $\xi\leq 0.555717769$, $KK^\dagger=\1_2$ in the ground state. Therefore, up to unitary rotations $K\to KU$ with $U\in \U(2)$, we have $K=\1_2$, i.e., $\langle K_{{\rm R}\up} \rangle=\langle K_{{\rm L}\down} \rangle =1$ and $\langle K_{{\rm R}\down} \rangle=\langle K_{{\rm L}\up} \rangle=0$. Hence the symmetry breaking $\U(1)_{\rm V}\times\U(1)_{\rm Q} \to \U(1)_{\rm V+Q}$ occurs. Furthermore, recalling that $K$ transforms under $\U(1)_{\rm A}\times\SU(2)_{\rm H}$ as 
\ba
	K \to \rme^{i\theta \sigma_3} K u \quad \text{with}~~\rme^{i\theta \sigma_3}\in\U(1)_{\rm A}~~\text{and}~~u\in\SU(2)_{\rm H}\,,
\ea
it follows that $K=\1_2$ spontaneously breaks $\U(1)_{\rm A}\times\SU(2)_{\rm H}$ down to $\U(1)_{\rm A+H}$. This is the random matrix incarnation of the chiral-HQS locking advocated in Refs.~\refcite{Yasui:2017izi,Yasui:2016svc}. 

In the coexistence phase at $0.555717769\leq \xi \leq 1$, both the Kondo condensate and the chiral condensate are nonzero. Analytically one can show
\ba
	a & = \frac{4-\xi^2-\xi\sqrt{\xi^2+8}}{8}\,,
	\label{eq:a_aesfd}
	\\
	\sigma & = \frac{\xi+\sqrt{\xi^2+8}}{4}\,.
	\label{eq:s_w23}
\ea
Specifically, we have $KK^\dagger=a\begin{pmatrix}1 & -1 \\ -1 & 1\end{pmatrix}$. There are infinitely many $K$ that fulfills this form due to unitary invariance ($K\to KU$). Let us mention two particular choices. The first one is $\langle K \rangle=\sqrt{a}\begin{pmatrix}1 & 0 \\ -1 & 0\end{pmatrix}$. Namely,
\ba
	\langle K_{{\rm R}\up}\rangle = -\langle K_{{\rm L}\up}\rangle = \sqrt{a} \quad \text{and} \quad 
	\langle K_{{\rm R}\down}\rangle=\langle K_{{\rm L}\down}\rangle=0\,,
\ea
implying that only up spin component of $Q$ participates in the Kondo effect. The second one is $\displaystyle \langle K \rangle=\sqrt{\frac{a}{2}}\begin{pmatrix}1 & 1 \\ -1 & -1 \end{pmatrix}$. This is a more symmetric choice, since all components equally contribute to the Kondo effect:
\ba
	\langle K_{{\rm R}\up}\rangle = \langle K_{{\rm R}\down}\rangle = -\langle K_{{\rm L}\up}\rangle 
	= -\langle K_{{\rm L}\down}\rangle=\sqrt{\frac{a}{2}}\,. 
	\label{eq:awerfr}
\ea
The relative minus sign of $\langle K_{\rm R}\rangle$ and $\langle K_{\rm L}\rangle$ in \eqref{eq:awerfr} arises due to the assumption that $\sigma\geq 0$. For either choice of $K$, one can conclude that a nonzero chiral condensate leads to breakdown of the chiral-HQS locking. 

Finally, the phase for $\xi\geq 1$ is an ordinary chiral symmetry broken phase with no Kondo effect. The only spontaneously broken symmetry is $\U(1)_{\rm A}$.

\section{\label{sc:eft}Low-energy effective theory}

The procedure to derive low-energy effective theory of soft fluctuations in RMT is established (see e.g., Refs.~\refcite{Shuryak:1992pi,Halasz:1995qb}). Following this route, in this section we examine the large-$N$ behavior of the partition function in the presence of external fields. In addition to quark masses, we shall insert the source term for the Kondo condensates
\ba
	j_{\rm R} \bar\psi_{\rm R} Q_\up + j_{\rm L} \bar\psi_{\rm L} Q_\down + j_{\rm R}^* \bar{Q}_\up \psi_{\rm R} + j_{\rm L}^* \bar{Q}_\down \psi_{\rm L}\,.
\ea
Inserting this into \eqref{eq:ZRMT} we find the fermion determinant (for $N_f=1$) to be
\ba
	& \det \begin{pmatrix}
	W+V & m & j_{\rm R} & 0
	\\
	m^* & -W^\dagger + V & 0 & j_{\rm L}
	\\
	j^*_{\rm R} & 0 & -V^\dagger+\lambda & 0
	\\
	0 & j_{\rm L}^* & 0 & -V^\dagger+\lambda
	\end{pmatrix}
	\notag
	\\
	& = \begin{cases}
	\det \big[(W+V)(W^\dagger-V)+|m|^2 \big]\;{\det}^2(-V^\dagger+\lambda)
	\\
	\qquad \text{for}~~j_{\rm L/R}\to 0,~\text{and}
	\\
	\det[(W+V)(V^\dagger-\lambda)+|j_{\rm R}|^2] \;
	\det[(-W^\dagger+V)(V^\dagger-\lambda)+|j_{\rm L}|^2]
	\\
	\qquad \text{for}~~m\to 0.
	\end{cases}
\ea
The sources $m$ and $j_{\rm L/R}$ play a role similar to a magnetic field applied to ferromagnets. Generally, in a phase with spontaneously broken symmetry, an infinitesimal external perturbation determines the orientation of a macroscopically large condensate. In the QCD vacuum, the quark mass, however small it is, determines the orientation of the chiral condensate. Recalling that the fermion determinant in QCD has the form $\det(\slashed{D}+m)$, such a phenomenon is possible only if $\slashed{D}$ has a macroscopically large number of eigenvalues in the vicinity of zero. This is the content of the Banks-Casher relation \cite{Banks:1979yr}. In our matrix model, $m$ couples to the matrix $(W+V)(W^\dagger-V)$ and $j$ to $(W+V)(V^\dagger-\lambda)$ and $(-W^\dagger+V)(V^\dagger-\lambda)$. Therefore, in a phase with nonzero chiral/Kondo condensates, these matrices must have a macroscopically large number of eigenvalues near the origin. The level fluctuations on the scale of a mean level spacing in this domain are believed to be universal \cite{Guhr:1997ve,Verbaarschot:2000dy}. 

One way to probe this regime in RMT is to directly diagonalize these matrices and derive spectral functions analytically. This would allow us to characterize the QCD Kondo phase through the microscopic operator spectrum. However, this direction seems difficult, if not impossible, for the current matrix model because $W$ and $V$ are intertwined in a peculiar way. Therefore we shall turn to another method: to derive the effective theory of the partition function coupled to external fields. It uncovers fruitful information about statistical properties of the underlying operator coupled to an external field \cite{Leutwyler:1992yt}. We emphasize that this is a nonperturbative analysis because the fluctuations of zero modes (i.e., spatially homogeneous modes) of NG fields are integrated out \emph{exactly.}

Our point of departure is \eqref{eq:rq354ewd} with source terms included and $\sigma$ rescaled ($\sigma\to \sigma/\sqrt{\eta}$):
\ba
	Z & \propto \int_\CC \rmd \sigma \int_{\CC^4}\rmd K\; 
	\rme^{- N |\sigma|^2 -N\Tr(K^\dagger K)}
	{\det}^N
	\left(\begin{array}{c|c}
	\begin{matrix}0&\sigma+\sqrt{\eta}m\\\sigma^*+\sqrt{\eta}m^*&0\end{matrix}& K + J
	\\\hline 
	K^\dagger + J^\dagger & \xi \1_2
	\end{array}\right)
	\label{eq:23842}
\ea
with $J\equiv \diag(j_{\rm R},j_{\rm L})$. We will take the large-$N$ limit with $m\sim j_{\rm L/R}\sim O(1/N)$ and define the microscopic variables
\ba
	\m \equiv Nm, ~~\j_{\rm L/R} \equiv N j_{\rm L/R}
	\quad \text{and} \quad 
	\J \equiv N J\,.
\ea

We begin with the pure chirally broken phase with $\langle\sigma\rangle=1$ and $\langle K \rangle=0$. Substituting $\sigma=\rme^{-i\phi}$ and $K=0$ into \eqref{eq:23842}, one can straightforwardly derive the large-$N$ behavior
\ba
	Z & \propto \int_0^{2\pi}\hspace{-2mm}\rmd \phi~\exp\kkakko{\sqrt{\eta}\mkakko{\m \rme^{i\phi} + \m^*\rme^{-i\phi}}}
	\label{eq:Z24378}
	\\
	& \propto I_0(2\sqrt{\eta}|\m|)\,.\label{eq:stpb}
\ea
Here $\phi$ is the NG mode of $\U(1)_{\rm A}$, and $I_0$ is the modified Bessel function of the first kind. The light and heavy quarks decouple. 

Next, we consider the pure Kondo phase in which $\langle \sigma \rangle=0$ and $\langle K \rangle=\1_2$. Substituting $\sigma=0$ and $K=U$ with $U\in\U(2)$ into \eqref{eq:23842} we readily obtain
\ba
	Z \propto \int_{\U(2)}\hspace{-3mm}\rmd U~\exp\big[\Tr(\J U^\dagger + U \J^\dagger)\big].
\ea
Interestingly enough, this coincides exactly with the finite-volume partition function of two-flavor QCD in the $\epsilon$-regime \cite{Leutwyler:1992yt,Shuryak:1992pi}. The role of the quark mass matrix is now played by $\J$. This integral can be done analytically \cite{Jackson:1996jb,Balantekin:2000vn} and yields (for real $\J$)
\ba
	Z &\propto \frac{\j_{\rm L} I_0(2\j_{\rm R})I_1(2\j_{\rm L}) - \j_{\rm R} I_0(2\j_{\rm L})I_1(2\j_{\rm R})}{\j_{\rm L}^2-\j_{\rm R}^2}\,.
	\label{eq:vow8434pg}
\ea

Finally we turn to the coexistence phase. This is by far the most complicated case. We denote the ground state as $\sigma_0\equiv \langle\sigma\rangle$ and $K_0\equiv \langle K \rangle$ [cf.~\eqref{eq:a_aesfd} and \eqref{eq:s_w23}]. The five NG modes can be introduced as
\ba
	\sigma & = \rme^{2i\phi}\sigma_0\,,
	\\
	K & = \rme^{i\phi\sigma_3}K_0 U \,, 
\ea
with $U\in\U(2)$. From \eqref{eq:23842} we then get
\ba
	Z & \propto \oint \rmd \phi \int_{\U(2)}\hspace{-3mm}\rmd U~{\det}^N
	\left(\begin{array}{c|c}
	\begin{matrix}0&\rme^{2i\phi}\sigma_0+\sqrt{\eta}m\\
	\rme^{-2i\phi}\sigma_0+\sqrt{\eta}m^*&0\end{matrix}& \rme^{i\phi\sigma_3}K_0 U + J
	\\\hline 
	U^\dagger K_0^\dagger \rme^{-i\phi\sigma_3} + J^\dagger & \xi \1_2
	\end{array}\right)
	\\
	& \approx \oint \rmd \phi \int_{\U(2)}
	\hspace{-3mm}\rmd U~\exp
	\bigg\{
		- \sqrt{\eta}\,\xi \Tr \left[R^{-1}
		\begin{pmatrix}0&\m \rme^{-2i\phi}
		\\
		\m^* \rme^{2i\phi}&0\end{pmatrix}
		\right]
	\notag
	\\
	& \qquad + 2 \; \mathrm{Re}\Tr\mkakko{
		R^{-1}\J \rme^{-i\phi\sigma_3}U^\dagger K_0^\dagger
	}
	\bigg\},
\ea
where
\ba
	R \equiv K_0 K_0^\dagger - \xi \sigma_0 \begin{pmatrix}0&1\\1&0\end{pmatrix}\,.
\ea
Notice that in the second term of the exponent, $\rme^{-i\phi\sigma_3}$ can be absorbed into $U$. It leads to factorization of the partition function. Using $K_0=\sqrt{a}\begin{pmatrix}1&0\\-1&0\end{pmatrix}$, we obtain
\ba
	Z & =\oint \rmd \phi~\exp\kkakko{\sqrt{\eta}\,C_1(\xi)(\m \rme^{-2i\phi}+\m^*\rme^{2i\phi})}
	\notag
	\\
	& \quad \times 
	\int_{\U(2)}\hspace{-3mm}\rmd U~\exp\ckakko{
		C_2(\xi) \,\mathrm{Re}\Tr\kkakko{\begin{pmatrix}\j_{\rm R} & -\j_{\rm L} \\ 0&0\end{pmatrix}U^\dagger}
	} 
	\label{eq:ytreo}
\ea
where
\ba
	C_1(\xi) & \equiv \frac{\xi+\sqrt{\xi^2+8}}{4}\,,
	\\
	C_2(\xi) & \equiv \sqrt{\frac{4-\xi^2-\xi\sqrt{\xi^2+8}}{2}} \,.
\ea
As a quick consistency check, notice that $C_1\to 1$ and $C_2\to 0$ as $\xi\to 1$, ensuring that the effective theory for the pure chirally broken phase \eqref{eq:Z24378} is smoothly recovered at the phase transition $\xi=1$.

The integrals \eqref{eq:ytreo} can be carried out exactly \cite{Jackson:1996jb,Balantekin:2000vn} and yields
\ba
	Z & \propto 
	I_0 \big( 2\sqrt{\eta}\,C_1(\xi)|\m| \big)
	\frac{
	I_1 \Big(
		C_2(\xi) \sqrt{|\j_{\rm R}|^2+|\j_{\rm L}|^2}
	\Big)}{\sqrt{|\j_{\rm R}|^2+|\j_{\rm L}|^2}}\,.
	\label{eq:9gdfc5wer3}
\ea
Analytical expressions \eqref{eq:stpb}, \eqref{eq:vow8434pg}, and \eqref{eq:9gdfc5wer3} encapsulate contributions of the zero modes of the NG bosons to the finite-volume QCD partition function to all orders in the loop expansion \cite{Gasser:1987ah,Leutwyler:1992yt,Shuryak:1992pi}. This is known as the $\varepsilon$-regime, where the thermodynamic limit is taken simultaneously with the vanishing limit of symmetry-breaking external sources. If we make sources $\m$ and $\j_{\rm L/R}$ much greater than $1$, we would gradually transition to the $p$-regime, where fluctuations of soft modes are suppressed.

\section{\label{sc:ccp}Chiral chemical potential}

\begin{figure}[tb]
	\centering
	\includegraphics[width=0.5\columnwidth]{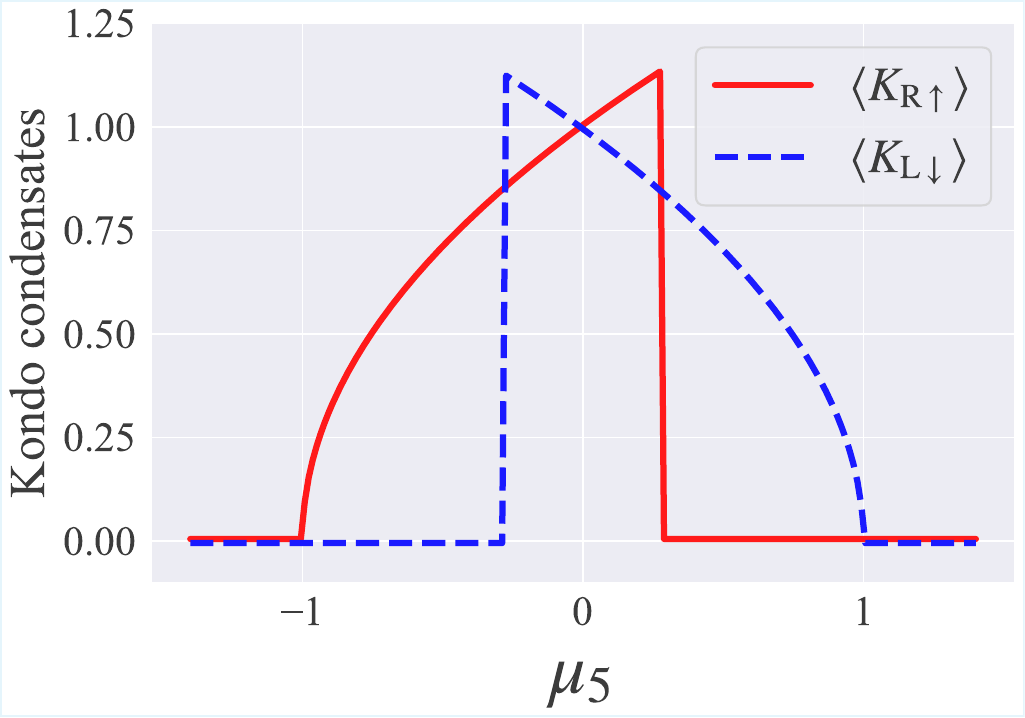}
	\vspace{-3mm}
	\caption{\label{fg:2wesfd}\revision{%
	The Kondo condensates $K_{{\rm R}\uparrow}$ and $K_{{\rm L}\downarrow}$ are plotted as a function of the chiral chemical potential $\mu_5$. Here, the subscripts R/L and $\uparrow/\downarrow$ denote the chirality of the light quark and the spin component of the heavy quark, respectively.}}
\end{figure}

The chiral chemical potential $\mu_5\bar\psi\gamma_0\gamma_5\psi$ introduces a net imbalance of chiralities. Its use is motivated by the chiral magnetic effect \cite{Kharzeev:2007jp,Fukushima:2008xe} in which a current is induced by a magnetic field in a chirally asymmetric matter. The interplay of the chiral chemical potential and the QCD Kondo effect was analyzed in an NJL-type model in Ref.~\refcite{Suenaga:2019jqu}. Here we discuss it in the context of our RMT. For simplicity, we switch off the interaction part of $W$, which amounts to neglecting chiral symmetry breaking in the light quark sector. By slightly modifying \eqref{eq:rq354ewd}, we get the partition function 
\ba
	Z & = \int \rmd\bar\psi\,	\rmd \psi\,\rmd \bar{Q}\, \rmd Q 
	\int_{\CC^4}\rmd K \rme^{-N\Tr(K^\dagger K)}
	\exp\left[
	\begin{pmatrix}\bar\psi_{\rm R}\\\bar\psi_{\rm L}\\\bar{Q}_\up\\\bar{Q}_\down\end{pmatrix}_\alpha
	\hspace{-4pt}
	\left(\begin{array}{c|c}
	\begin{matrix}\rho&0\\0&-\rho\end{matrix}& K
	\\\hline 
	K^\dagger &\begin{matrix}\lambda&0\\0&\lambda\end{matrix}
	\end{array}\right)
	\begin{pmatrix}\psi_{\rm R}\\\psi_{\rm L}\\Q_\up\\Q_\down\end{pmatrix}_\alpha
	\right]
\ea
where $\rho$ represents the chiral chemical potential. Integrating out fermions yields
\ba
	Z & \propto \int_{\CC^4}\rmd K \ckakko{\rme^{-\Tr(K^\dagger K)}\det(KK^\dagger-\mu_5\sigma_3)}^N
\ea
with 
\ba
	\mu_5 \equiv \rho \lambda\,.
\ea
Again we employ the parametrization \eqref{eq:KKp}. As $\det(KK^\dagger-\mu_5\sigma_3)$ depends on $c$ and $d$ through the combination $c^2+d^2$ one can set $d=0$ without loss of generality. The task is to minimize the free energy
\ba
	f_5(a,b,c) \equiv a + b - \log \left| (a-\mu_5)(b+\mu_5) - c^2 \right|
\ea
under the conditions $a\geq 0, b\geq 0$ and $ab\geq c^2$. We solved this problem numerically and found that $c=0$ for all $\mu_5$. The condensates are then obtained as $\langle K_{{\rm R}\up}\rangle=\sqrt{a}$ and $\langle K_{{\rm L}\down} \rangle=\sqrt{b}$. The numerical results are plotted in Figure~\ref{fg:2wesfd} and are summarized below.
\begin{center}
	\begin{tabular}{@{}ccc@{}}
		\toprule
		Domain & $\langle K_{{\rm R}\up} \rangle$ & $\langle K_{{\rm L}\down} \rangle$ \\
		\midrule\midrule
		$-1 \le \mu_5 \le -\mu_5^{C}$ & $\sqrt{1+\mu_5}$ & $0$ \\
		$-\mu_5^{C} \le \mu_5 \le \mu_5^{C}$ & $\sqrt{1+\mu_5}$ & $\sqrt{1-\mu_5}$ \\
		$\mu_5^{C} \le \mu_5 \le 1$ & $0$ & $\sqrt{1-\mu_5}$ \\
		$|\mu_5|>1$ & $0$ & $0$ \\
		\bottomrule
	\end{tabular}
\end{center}
There are phase transitions at $\mu_5=\pm 1$ and $\pm \mu_5^C$, where
\ba
	\mu_5^C \equiv 0.27846\cdots 
\ea
is the solution of $\rme^{1+x}x=1$. 
As expected on physical grounds, $\mu_5$ lifts the degeneracy of left-handed and right-handed Kondo condensates. At large $|\mu_5|$, the condensates go away, which we think is an artifact of the model, similar to the UV cutoff effects in NJL models.

\section{\label{sc:conc}Conclusions and outlook}

In this work, we have constructed a new random matrix model that captures the essential aspects of the QCD Kondo effect. The model has correct symmetries of QCD with a heavy flavor. We solved the model for $N_f=1$ in the large-$N$ limit with $N$ the matrix size, and classified the patterns of spontaneous symmetry breaking. The model has three phases that arise due to the competition of chiral symmetry breaking and the QCD Kondo effect. In the Kondo phase with no chiral condensate, we have shown that the $\U(1)_{\rm A}$ symmetry of light quarks and the $\SU(2)$ heavy quark spin symmetry are locked to the diagonal $\U(1)$ subgroup, as has been suggested in preceding work \cite{Yasui:2017izi,Yasui:2016svc}. In the phase where the Kondo effect and chiral symmetry breaking coexist, we have shown that the pairing form of the Kondo condensate is drastically modified. This is an important prediction of our model. Moreover, for each of the three phases we derived the low-energy effective theory of NG modes and obtained closed formulas for the partition function. We hope this work contributes to better analytical understanding of the QCD phase diagram with heavy quarks. 

One caveat is that our model is dimensionless and therefore cannot predict quantities in physical units. For example, determining the location of a phase transition in units of $\Lambda_{\rm QCD}$ or in GeV would require a more microscopic framework with propagating degrees of freedom. Nevertheless, we expect that the qualitative insights our model provides into the competition between orders in the ground state will be useful for future, more microscopic studies.

There are various future directions of research. 
\begin{itemize}
	\item The analysis in this paper is limited to $N_f=1$. Extension to $N_f>1$ is a challenging open problem.
	\item The axial anomaly was ignored in this work. How to incorporate anomaly into our random matrix model is not clear yet.
	\item Our prediction for the pairing form of the Kondo condensate in the coexistence phase should be checked in more realistic (e.g., NJL-type) models.
	\item As is well known, there are three symmetry classes in chiral RMT, corresponding to quarks in a complex/real/pseudoreal representation of the gauge group \cite{Verbaarschot:1994qf}. The matrix model in this work corresponds to QCD with quarks in a complex representation. Extension to the other two classes (the chiral orthogonal ensemble and the chiral symplectic ensemble) will be very interesting.
\end{itemize}

\bibliography{manuscript.bbl}
\end{document}